\documentclass[
reprint,
twocolumn,
amsmath, 
amssymb,
aps, 
prl,
]{revtex4-2}

\usepackage{graphicx} 
\usepackage{dcolumn}  
\usepackage{bm}       
\usepackage{calc}
\usepackage{physics}
\usepackage{bbm}
\usepackage{verbatim}
\usepackage[dvipsnames]{xcolor}
\usepackage[colorlinks, citecolor=blue, linkcolor=blue]{hyperref}
\usepackage[section]{placeins}

\newcommand{\ddo}{ö}

\begin{document}

\title{Test of Causal Non-Linear Quantum Mechanics by Ramsey Interferometry on the vibrational mode of a trapped ion}

\author{Joseph Broz}
\affiliation{
 Department of Physics, University of California, Berkeley, California 94720, USA
}
\affiliation{Challenge Institute for Quantum Computation, University of California, Berkeley, CA  94720}
\author{Bingran You}
\affiliation{
 Department of Physics, University of California, Berkeley, California 94720, USA
}
\affiliation{Challenge Institute for Quantum Computation, University of California, Berkeley, CA  94720}
\author{Sumanta Khan}
\affiliation{
 Department of Physics, University of California, Berkeley, California 94720, USA
}
\affiliation{Challenge Institute for Quantum Computation, University of California, Berkeley, CA  94720}
\author{Hartmut Häffner}
\affiliation{
 Department of Physics, University of California, Berkeley, California 94720, USA
}
\affiliation{Challenge Institute for Quantum Computation, University of California, Berkeley, CA  94720}

\author{David E. Kaplan and Surjeet Rajendran}
\affiliation{
 Department of Physics and Astronomy, The Johns Hopkins University, Baltimore, Maryland 21218, USA
}

\date{\today}

\begin{abstract}
Kaplan and Rajendran have recently demonstrated that non-linear and state-dependent terms can be consistently added to quantum field theory to  yield causal non-linear time evolution in quantum mechanics. Causal non-linear theories have the unavoidable feature that their quantum effects are dramatically sensitive to the full physical spread of the quantum state of the system. As a result, such theories are not well tested by conventional atomic and nuclear spectroscopy. By using a well-controlled superposition of vibrational modes of a $^{40}$Ca$^+$ ion trapped in a harmonic potential, we set a stringent limit of $5.4\times10^{-12}$ on the magnitude of the unitless scaling factor $\tilde{\epsilon}_{\gamma}$ of the predicted causal, non-linear perturbation.
\end{abstract}

\maketitle


\textcolor{purple}{\textit{Introduction.}}--A basic feature of quantum mechanics (QM) is that the time evolution of the wave function is described by a linear equation of motion. This is treated as an axiom in deriving many elementary results, like conservation of probability or the no-cloning theorem \cite{Wigner1939OnUR, Wootters1982Oct}. But, interestingly, there is no proof that linearity is necessary to obtain a physically consistent theory, raising the possibility that QM evolution could be non-linear.

This is evidenced by a number of successful non-linear generalizations of QM (NLQM) for {\it single} particles that have been independently fleshed out in the literature \cite{Bialynicki-Birula1976Sep, Weinberg1989Jan, Weinberg1989Sep, doebner-goldin}. These examples demonstrate that state-dependent terms can be added to the non-relativistic Schr\ddo dinger Hamiltonian while maintaining a compatible physical interpretation and correspondence with linear QM. The rigorous experimental tests that have followed \cite{Shull1980Mar, Gahler1981Apr, Bollinger1989, Chupp1990May, Walsworth1990May, Majumder1990Dec} further highlight the seriousness of these proposals and have also imposed stringent bounds on their predictions.

But QM needs to be able to describe multiple particles that can exist together in an arbitrarily complicated entangled state.  Naive generalizations of \cite{Bialynicki-Birula1976Sep, Weinberg1989Jan, Weinberg1989Sep, doebner-goldin} to such states generally lead to violations of causality \cite{GISIN1, Gisin2, Polchinski}, contributing to a widespread belief that linearity is necessary for causality \cite{Bassi2015Aug}. However, as first pointed out by Polchinski \cite{Polchinski}, causal NLQM evolution of multi-particle states is possible if the non-linear terms in the Schr\ddo dinger equation are restricted to a specific form.

Recently, Kaplan and Rajendran \cite{KaplanandRajendran}, building on earlier work by Kibble \cite{Kibble:1978vm}, have developed a systematic approach for incorporating causal non-linear evolution into quantum field theory (QFT). The introduction of non-linearities directly into QFT as opposed to the single particle Schr\ddo dinger equation is motivated by the fact that QFT is the natural framework to describe the causal evolution of multi-particle states. Excitingly, it was shown in \cite{KaplanandRajendran} that the non-linear structure demanded by \cite{Polchinski} for multi-particle states was a natural consequence of QFT. 

The basic approach of \cite{KaplanandRajendran} is to start with a given QFT and introduce non-linearities by shifting bosonic field operators by a small amount proportional to the expectation value of the field operator acting on the full quantum state. When applied to single particle systems, the procedure yields a non-linear Schrödinger equation. For example, the time evolution of a single particle with charge $q$ and Hamiltonian $H$ is described in this theory by:

\begin{align} 
    i\hbar\partial_t\Psi(t,\mathbf{x}) = &\bigg(H + \tilde{\epsilon}_{\gamma} \frac{q^2}{4\pi\varepsilon_0}\int d^4 \mathbf{x}_1 |\Psi(t_1,\mathbf{x}_1)|^2G_r(t, \mathbf{x}; t_1, \mathbf{x_1})\bigg)\nonumber\\
    &\times\Psi(t,\mathbf{x}) \label{relativisticcorrection}
\end{align}

\noindent where $\tilde{\epsilon}_{\gamma}$ is a small unitless parameter scaling the non-linearity of the theory \footnote{The $\gamma$ subscript specifies that this perturbation is specific to electromagnetic fields. The causal theory is field-dependent and does not explicitly require that the scaling of the perturbation be the same for self-interactions mediated by different quantum field theories.} and $G_r$ is the relativistic retarded Green's function from the spacetime coordinates $(t_1,\mathbf{x}_1)$ to $(t,\mathbf{x})$ . $G_r$ naturally appears in this expression from the underlying QFT derivation and enforces causality.  The new term added to the Hamiltonian in Eq. \eqref{relativisticcorrection} admits the simple interpretation of a classical Coulomb potential causally sourced by the quantum probability distribution of the particle's position. 

As with past efforts, Kaplan and Rajendran's theory also maintains a satisfactory correspondence with standard QM, preserving important features like conservation of probability and energy, the existence of stationary states and a consistent notion of measurement. One might expect, however, that the strong bounds set from previous searches for NLQM \cite{Shull1980Mar, Gahler1981Apr, Bollinger1989, Chupp1990May, Walsworth1990May, Majumder1990Dec} would largely carry over. But this expectation turns out to be false for the following fundamental reason. The tests performed by \cite{Shull1980Mar, Gahler1981Apr, Bollinger1989, Chupp1990May, Walsworth1990May, Majumder1990Dec} are on energy levels of various bound states. In linear QM, the level structure is independent of the center of mass spread of the bound state wave-function. This is not true in causal NLQM where non-linear effects alter time evolution via the position space wavefunction as in Eq.~\eqref{relativisticcorrection}. These effects are highly suppressed if the center of mass wave-function is spread out. 
To illustrate this point, it is helpful to take the non-relativistic limit of Eq. \eqref{relativisticcorrection}. When $||H||/\hbar \ll c/|\mathbf{x}_1-\mathbf{x}|$ the non-linear Schr\ddo dinger equation becomes:

\begin{equation} \label{nonrelativisticcorrection}
    i\hbar\partial_t\Psi(t,\mathbf{x}) = \bigg(H + \tilde{\epsilon}_{\gamma} \frac{q^2}{4\pi\varepsilon_0}\int d^3 \mathbf{x}_1 \frac{|\Psi(t,\mathbf{x}_1)|^2}{|\mathbf{x}_1 - \mathbf{x}|}\bigg) \Psi(t,\mathbf{x})
\end{equation}
Here one can see that denominator of the integrand scales with the full position-space spread of the wave function, damping the perturbation accordingly. This is a simple consequence of the Coulomb potential that sources the non-linearity, but the implication is that any sensitive test based on standard atomic or nuclear spectroscopy must also pin down the test system's center of mass motion to a dimension comparable to the spread of the internal degrees of freedom. This condition was not well-satisfied in previous tests for NLQM \cite{Shull1980Mar, Gahler1981Apr, Bollinger1989, Chupp1990May, Walsworth1990May, Majumder1990Dec}, but by requiring the non-linear correction to be smaller than the uncertainty in recent Lamb shift measurements of hydrogen, Kaplan and Rajendran have set a modest bound of $|\tilde{\epsilon}_{\gamma}| \lessapprox 10^{-2}$, giving a sense for the limitations of atomic spectroscopy \cite{KaplanandRajendran}.

For a more precise test, one might perform Ramsey spectroscopy \cite{Ramsey} on a superposition of the Fock states $|n\rangle$ of a harmonic vibrational mode of a trapped ion \cite{KaplanandRajendran}. The state $|\psi(t{=}0)\rangle=\alpha_n|n\rangle + \alpha_m|m\rangle$ can be prepared and then allowed to freely evolve for an interrogation time $\tau$. The Coulomb field sourced by the position-space expectation value of $|\psi\rangle$ interacts differently with the two branches of the wave function leading to an energy shift and thus the accumulation of a measurable phase difference between them \cite{Schmidt_Kaler_2003}. 
The advantage of this method is that a superposition can be created where 1) the physical spread of the center-of-mass wave function is well localized with respect to the size of the wavefunction and 2) there is still very little overlap between the position space distributions of the two branches. The first point ensures that non-linear perturbation is not small and the second point ensures that the effect it has on the two branches of the wave function is dissimilar -- maximizing the phase difference.

These conditions are satisfied when $n{=}0$ and $m{=}1$, i.e. the initial state is a superposition of the ground and first excited state. If one replaces $H$ in Eq. \eqref{nonrelativisticcorrection} with the Hamiltonian for a three-dimensional, isostropic harmonic oscillator \footnote{For an anisotropic potential, Eq. \eqref{phi} will incur an $\mathcal{O}(\tilde{\epsilon}_{\gamma})$ correction.} and assumes that the vibrational modes in the two transverse directions remain in their ground state, the phase difference accumulated between the ground and first excited state of the superposition after a time $\tau$ is given by:

\begin{equation} \label{phi}
    \phi_{NL}(\tau; \{\alpha_i\}) = \tilde{\epsilon}_{\gamma}\frac{10 \alpha_0^2 + \alpha_1^2}{30\sqrt{2\pi}\hbar}  \frac{e^2}{4\pi\varepsilon_0 x_0} \tau
\end{equation}

\noindent where the $\alpha_i$ are assumed to be real and $x_0 = \sqrt{m\nu/\hbar}$ is the characteristic length scale of a harmonic oscillator with mass $m$ and natural frequency $\nu$. Note that the state-dependence of $\phi_{NL}$, i.e. its dependence on the weight of the energy eigenstates via $\alpha_i$, is a characteristic non-linear effect, which has no analog in the linear theory.  For an ion localized to $x_0 = 10 $\,nm, a phase of up to order $10^{10}\times \tilde{\epsilon}_{\gamma}$ is accumulated for every millisecond of interrogation time. In this Letter we perform such a Ramsey experiment designed to maximize the signal $\phi_{NL}$ and thus tighten the bound on $\tilde{\epsilon}_{\gamma}$ by 8 orders of magnitude relative to the current best estimate.




\begin{figure*}[ht]
\includegraphics[width=0.99\linewidth]{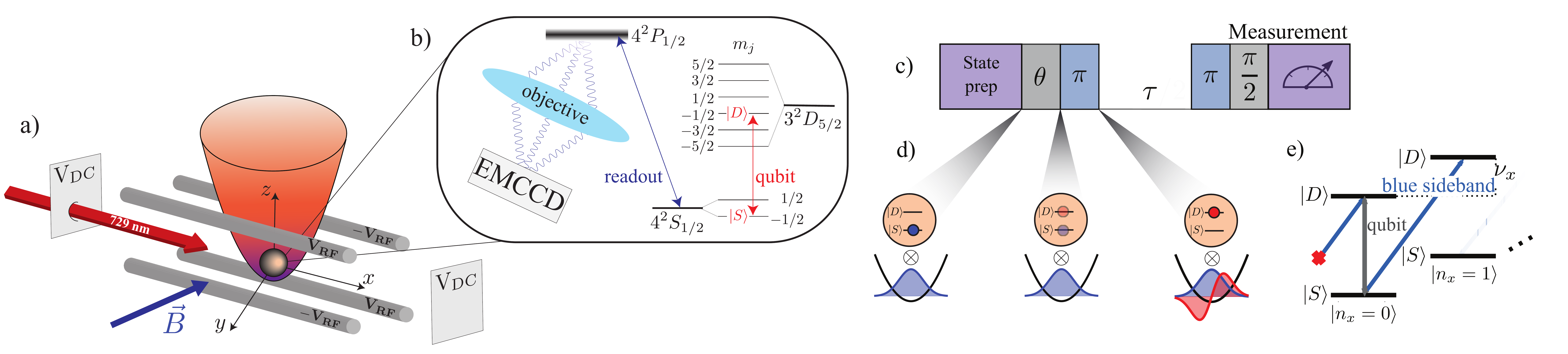}
\caption{\label{fig:experiment-illustration}
    \textbf{Experimental implementation.} \textbf{(a)} A $^{40}$Ca$^+$ ion is trapped using a combination of RF and DC electric fields. In a time-averaged sense, the confinement is well modeled by a 3-dimensional harmonic potential. \textbf{(b)} Motion along the $x$-direction is excited by resonantly coupling to the internal electronic Zeeman sublevels of the $4^2S_{1/2} \leftrightarrow 3^2D_{5/2}$ transition using narrow band light near 729\,nm. The degeneracy of the Zeeman states is broken through application of a strong magnetic field of $\approx$4\,G. Measurement is performed by scattering photons of the short-lived $4^2S_{1/2}\leftrightarrow 4^2P_{1/2}$, which are then focused onto an EMCCD camera. \textbf{(c)} The experimental pulse sequence. Pulses that address the qubit are colored gray and those that address the blue sideband, blue. After preparing the state $|S,0\rangle$, the first pair of pulses is used to generate the state $|\psi(t=0)\rangle = |D\rangle (\alpha_0|0\rangle + \alpha_1|1\rangle)$. This is then allowed to freely evolve for a time $\tau$, accumulating a relative phase of $\Phi(\tau)$, which is sensitive to the proposed causal non-linear perturbation. Afterwards, the information is mapped onto the qubit with a blue sideband pulse and then the expectation value of the Pauli spin operator $\text{cos}(\xi_L)\sigma_x + \text{sin}(\xi_L)\sigma_y$ is measured. \textbf{(d)} An illustration of the two-step process for generating the state $|\psi(t=0)\rangle$, as described in more detail in the main text. The key feature is the fact that the state $|D,0\rangle$ is transparent to the resonant blue sideband drive as illustrated in \textbf{(e)}. This allows us to map an arbitrary qubit state onto the ground and first excited state of the vibrational mode.
}
\end{figure*}

\textcolor{purple}{\textit{Experimental implementation.}}-- Experiments were performed using a single $^{40}$Ca$^+$ ion confined inside of a radio-frequency (RF) Paul trap in a parameter regime where the center of mass motion is well modelled as a 3-dimensional anisotropic harmonic oscillator (Fig. \ref{fig:experiment-illustration}a) with vibrational frequencies  $\nu_x \approx 2\pi \times 1.01$ MHz, $\nu_y \approx 2\pi \times 2.52$ MHz and $\nu_z \approx 2\pi \times 2.79$ MHz.

The ion's internal state is manipulated by shining resonant
laser light on various electronic transitions (Fig. \ref{fig:experiment-illustration}b). The short-lived $4^2S_{1/2}\leftrightarrow{4^2P_{1/2}}$ and $3^2D_{5/2}\leftrightarrow{4^2P_{3/2}}$ dipole transitions are used for entropy-altering operations like cooling and measurement. Measurement, in particular, is performed via the electron shelving method on $4^2S_{1/2}\leftrightarrow{4^2P_{1/2}}$ and allows us to determine the population of the $4^2S_{1/2}$ manifold \cite{electron_shelving}. For coherent operations, narrowband light at 729\,nm is used to couple the $|4^2S_{1/2},m_J{=}-1/2\rangle$ and $|{3^2D_{5/2}},m_J{=}-1/2\rangle$ states, whose degeneracy is broken with a static magnetic field of $B\approx$ 4\,G. We call this our qubit transition and reference it as $|S\rangle\leftrightarrow|D\rangle$.

To prepare the ion in a well-defined state, we first cool its temperature to several hundred microkelvin using Doppler cooling and then optically pump its electronic state into $|S\rangle$. Afterwards, resolved sideband cooling is applied along the $x$-direction, driving the axial vibrational mode into its ground state with high probability \cite{sidebandcooling}. Once this process is complete, the ion is measured to be in the state $|S, n_x=0\rangle$ with a confidence greater than $99\%$, where $n_x$ refers to the phonon number of the vibrational mode along the $x$-direction. The two transverse vibrational modes are left in thermal states with mean phonon occupations determined by the Doppler-limit of $\langle n_{y,z}\rangle\approx3$). These modes remain separated from the $|S, n_x\rangle$ state and so we ignore them in what follows except for taking into account the additional spread of the wave function in position space to determine the nonlinearity in Eqs.~\ref{nonrelativisticcorrection} and \ref{phi}.

In order to create the desired superposition state, we use laser light resonant with the a motional sideband of the qubit transition.
From the ion's perspective, a laser pointing along one of its vibrational axes will appear to be phase modulated by motion along that direction. By detuning the laser from the qubit frequency by an amount equal to $+\nu_x$, this effect can be used to couple the states ${|S,n\rangle\leftrightarrow|D,n+1\rangle}$, which we refer to as blue sideband transitions (Fig. \ref{fig:experiment-illustration}e) \cite{Leibfried2003Mar}. The energy of the blue sideband transitions is already sensitive to the non-linear perturbation and, in principle, can be used for our Ramsey experiment. But the electronic states are first-order sensitive to ambient magnetic field fluctuations leading to a coherence time an order of magnitude less than that of the vibrational mode -- unnecessarily limiting the Ramsey interrogation time.

So, instead we first map the desired Ramsey superposition onto the ion's internal states by resonantly driving the qubit transition for a fixed duration, generating the state $(\alpha_1 |S\rangle +  \alpha_0 |D\rangle)|0\rangle$, in an appropriate rotating frame. Here $\alpha_0 = \text{sin}(\theta/2)$, $\alpha_1 = \text{cos}(\theta/2)$ and the value of $\theta$ is controlled by adjusting the intensity of the addressing laser. Next, we drive a blue sideband $\pi$-pulse that nominally transfers all of the population from $|S,0\rangle$ to $|D,1\rangle$ but leaves the population in $|D,0\rangle$ untouched (Fig. \ref{fig:experiment-illustration}d-e). Together, these operatons result in the separated state ${|\psi(t=0)\rangle = |D\rangle(\alpha_0|0\rangle + \alpha_1|1\rangle)}$, where the qubit state information has been written onto the vibrational mode \cite{Schmidt_Kaler_2003}.

Once the state $|\psi\rangle$ has been prepared, it is allowed to evolve freely for a time $\tau$ so that a relative phase $\Phi(\tau;\theta)$ is accumulated and ${|\psi(\tau)\rangle = |D\rangle(\alpha_0|0\rangle + e^{i\Phi(\tau,\theta)}\alpha_1|1\rangle)}$. To extract this phase, we repeat the steps used to generate $|\psi(0)\rangle$ in a time-reversed order (with the value of $\theta$ now fixed at $\pi/2$ where the signal is maximized) and then measure $|D\rangle$, which will be occupied with a probability of:

\begin{equation} \label{ramseysignal}
    P(\tau) = B - \frac{A(\tau)}{2}\;\text{cos}[\Phi(\tau;\theta) + \xi_L]
\end{equation}

\noindent here $0\le A(\tau)\le 1$ is the signal contrast which will generally be less than one when $\theta\ne\pi/2$, $B\approx1/2$ is an offset whose precise value is sensitive to errors in state preparation/ measurement and $\xi_L$ is the laser phase of the final qubit $\pi/2$-pulse relative to the initial $\theta$-pulse. Since $P(\tau)$ is an expectation value, a single estimate is obtained by repeating the experiment 200 times, which is large enough that the propagation of the quantum projection noise (QPN) converges when inverting Eq. \eqref{ramseysignal}. The full pulse sequence is illustrated in Fig. \ref{fig:experiment-illustration}c. 

\begin{figure}
\includegraphics[width=0.99\linewidth]{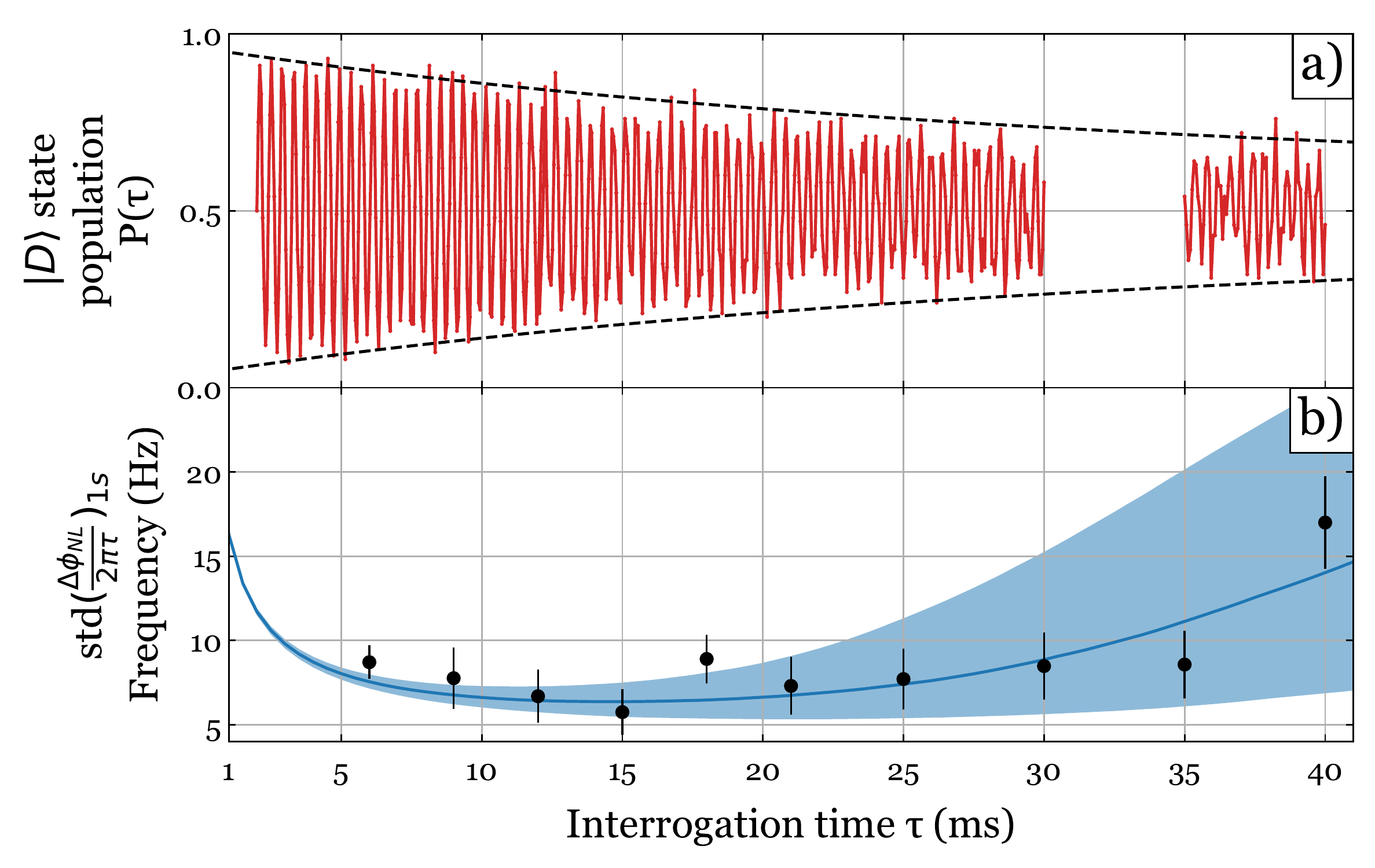}
\caption{\label{fig:optimal_time}
    \textbf{Experimental Performance.} \textbf{(a)} Measured $P(\tau)$, as described by Eq. \eqref{ramseysignal} (red). The black dashed line is the predicted decay envelope taking into account only heating of the vibrational mode at a rate of 10 quanta/s. The reasonable agreement between the predicted and measured decay suggests that the Ramsey signal contrast is dominated by this heating process. \textbf{(b)} The black circles represent the sample standard deviation from repeated measurements of $\Delta\phi_{NL}(\tau)$ taken at various interrogation times and normalized to an integration time of 1\,s. The blue shaded region bounds the simulated predictions assuming only QPN and a heating rate between 7 and 13 quanta/s (lower and upper edge of the region, respectively). The dark blue line corresponds to 10 quanta/s.
}
\end{figure}

In order to gauge the performance of the Ramsey experiment, we conduct a test experiment where we apply a detuning $\Delta$ from resonance of several kHz to the first blue sideband pulse. In the rotating frame, this breaks the degeneracy of $|0\rangle$ and $|1\rangle$ leading to a phase of $\Phi(\tau) = \Delta\tau$ and, thus, sinusoidal oscillations of $P(\tau)$. The result is plotted in Fig. \ref{fig:optimal_time}a, where one can see that the signal contrast $A(\tau)$ exhibits a clear time-dependence due to zero-mean noise effects beyond the simple model described in Eq. \eqref{ramseysignal}. 

The dominant source of this noise is found to be a Markovian heating of the vibrational mode caused by ambient electric field fluctuations at the position of the ion and, perhaps, high frequency noise on the trapping potential \cite{Schneider1999May}. This means that during free evolution, the vibrational mode may spontaneously absorb a phonon from its environment with a probability that grows linearly in time. When $n$ phonons are absorbed, the state of the system after the final blue sideband pulse will be $|S\rangle (\alpha_0|n\rangle + \alpha_1|n+1\rangle)$ and the result of the final $\pi/2$-pulse, regardless of $\Phi$, will be a symmetric distribution of $\{|S\rangle,\;|D\rangle\}$ -- diminishing the averaged signal contrast. The dashed line in Fig. \ref{fig:optimal_time}a is a simulated decay envelope computed assuming only this heating process as characterized by the heating rate $\dot{\bar{n}}\approx10$\, quanta/s, independently measured by monitoring the red sideband \cite{Leibfried2003Mar}. The good agreement between the simulated and measured decay validates our earlier claim that the contrast is limited by environmental heating.

For a precise determination of the non-linearity it is most convenient to estimate $\Phi(\tau)$ at a fixed $\tau$. But since $A$, $B$ and $\Phi$ are all empirical quantities, inverting Eq. \eqref{ramseysignal} requires at least three independent measurements. We obtain these by repeating the experiment for three different values of $\xi_L$ spaced by ninety degrees such that $\xi^{(3)}_L = \xi^{(2)}_L + \pi/2 = \xi^{(1)}_L + \pi$. The targeted value of $\xi^{(1)}_L$ is chosen to minimize the standard deviation of $\Phi(\tau)$: 

\begin{equation} \label{stdphi}
    |\delta \Phi(\tau)|=\sqrt{\sum_{i}\left(\frac{\partial \Phi}{\partial \mathrm{P}_{i}} \delta \mathrm{P}_{i}\right)^{2}}
\end{equation}

\noindent which occurs when $\Phi(\tau) + \xi^{(1)}_L = \pi/2$. Here $P_i$ is the population measurement associated with $\xi_L^{(i)}$ and $\delta P_i$ is its standard deviation, nominally dominated by QPN.

A single measurement of $\Phi(\tau)$ contains the non-linear signature $\phi_{NL}(\tau)$, as described by Eq. \eqref{phi}, but also includes information about the detuning $\Delta$ of the blue sideband pulses from resonance and any AC Stark shifts that occur during state preparation and readout. Explicitly: $\Phi(\tau;\theta) = \phi_{NL}(\tau;\theta) + \Delta\tau + \phi_{SS}$, where $\phi_{SS}$ is the phase imprinted by the Stark shifts. Ideally, the frequency of the blue sideband pulses are calibrated such that $\Delta=0$, but slow drifts of the trapping potential on a time scale that is long relative to the Ramsey interrogation time generally cause this condition to be violated.
Likewise, Stark shifts incurred while driving the blue sideband cause a phase offset. But importantly, both $\Delta$ and $\phi_{SS}$ are independent of $\theta$ meaning that we can obtain an unbiased estimate of the non-linearity by repeating the measurement for two different values of $\theta$ and taking their difference:

\begin{align} \label{deltaphinl}
    \Delta\phi_{NL}(\tau; \{\theta_i\}) &= \Phi(\tau; \theta_1) - \Phi(\tau; \theta_2) \nonumber\\
     &= \phi_{NL}(\tau; \theta_1) - \phi_{NL}(\tau;\theta_2)
\end{align}

\noindent We choose $\theta_1$ and $\theta_2$ such that the ground state population of $|\psi(t=0)\rangle$ is 0.2 and 0.8, respectively. We also verify that there is not phase difference due to the Stark shift for both preparation sequences.

The non-linear signal $\Delta\phi_{NL}$ grows linearly with interrogation time $\tau$. But this effect must contend with the contrast decay and QPN, both of which increase the uncertainty of the signal (Eq. \eqref{stdphi}) and both of which favor shorter, more frequent measurements \cite{Huelga1997Nov}. The combination of these effects results in an optimal interrogation time, which we determine experimentally by measuring $\Delta\phi_{NL}(\tau)$ at various $\tau$ and computing the sample standard deviation. These results are normalized to an integration time of 1\,s and plotted in Fig. \ref{fig:optimal_time}b. The blue shaded region is a corresponding simulation that assumes only QPN and vibrational heating bounded by $7\le\dot{\bar{n}}\le13$\,quanta/s. Based on this data, we fix $\tau = 15$\,ms.

\begin{figure}
\includegraphics[width=0.99\linewidth]{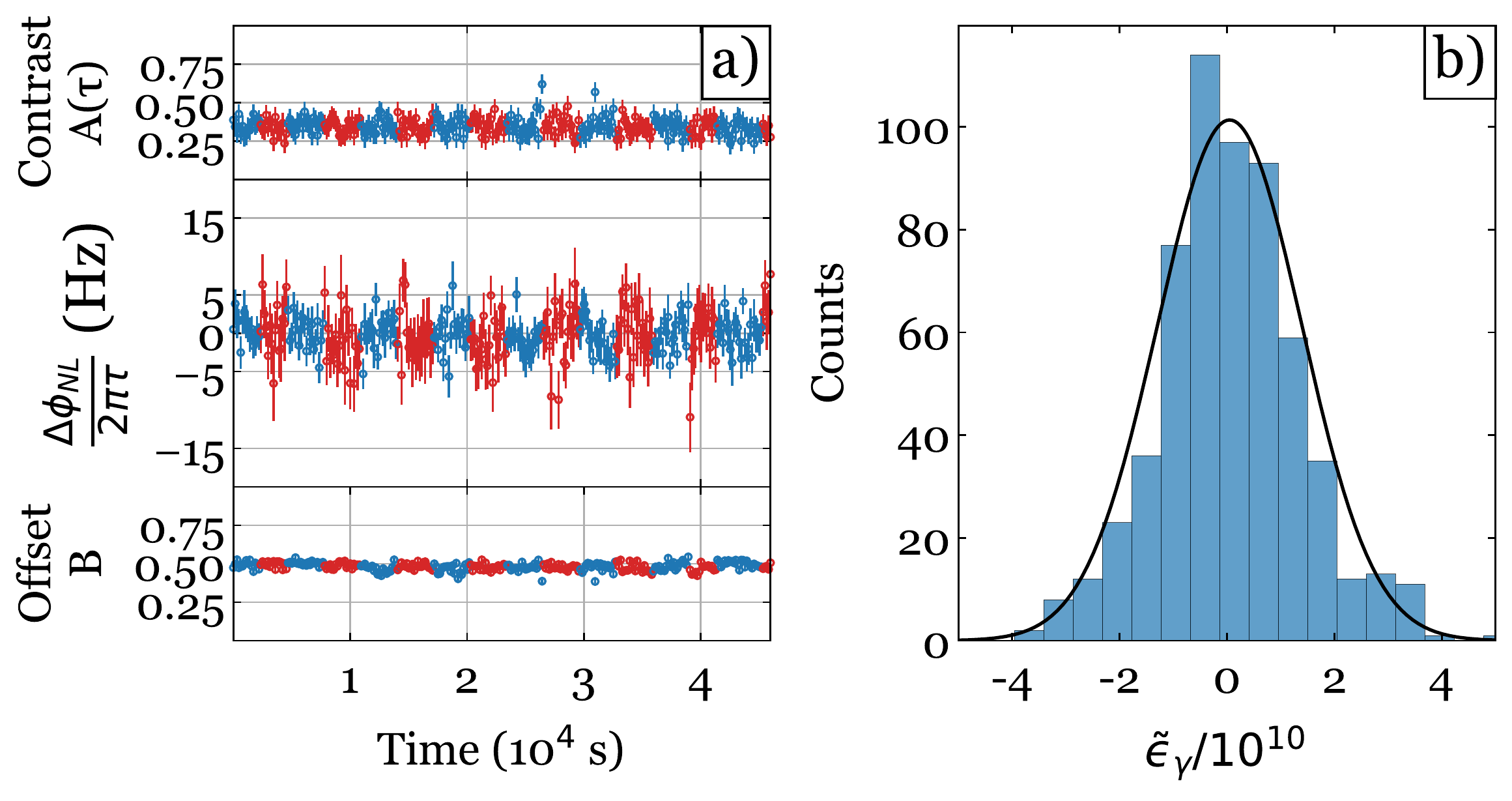}
\caption{\label{fig:results}
    \textbf{Results.} \textbf{(a)} (top to bottom) The measured contrast, frequency and offset over a full day of data collection. The blue circles represent data taken at an interrogation time of 15\,ms and the red circles were taken at a time of 15\,ms divided by the golden ratio ($\approx$9.3\,ms). \textbf{(b)} The distribution of $\tilde{\epsilon}_{\gamma}$ estimated from the data. The mean value is $5 \pm 5.4 \times 10^{-12}$. The black curve is a Gaussian fit to the distribution.
}
\end{figure}

To determine a more rigorous bound on $\tilde{\epsilon}_{\gamma}$ we repeat the measurement of $\Delta\phi_{NL}(\tau{=}15\,\text{ ms})$ many times. Before each $\phi_{NL}$ measurement, we independently measure the initial qubit excitation to determine the precise values of $\theta_i$ which may change slightly over time due to intensity drifts of the addressing light. We also perform a preliminary 3-point Ramsey measurement with the population of $|0\rangle$ set to 0.5 to produce a maximum signal that we use to optimally bias $\xi^{(1)}_L$. Next, we perform 1200 measurements of the Ramsey signal, 200 for each of the 3 Ramsey points for $\theta_1$ and $\theta_2$. The ordering of these experiments is randomized so as to avoid a bias due to drifts in $\Phi(\tau)$. From this data we compute $\Delta\phi_{NL}$, the average contrast $A$  of the two runs and the average offset $B$. For a single day of data, this is plotted in Fig. \ref{fig:results}a. The blue dots show data taken at $\tau=15$ ms. The red dots show data taken at 15\,ms divided by the golden ratio $(1+\sqrt{5})/2 \approx 9.27$\,ms, which does not  improve the estimate of the non-linearity but allows us to rule out the remote possibility that $\Delta\phi_{NL}(\tau{=}15\,\text{ ms})$ modulo $2\pi$ vanishes even though the perturbation is not small.

The distribution of $\tilde{\epsilon}_{\gamma}$ computed from the measured values of $\Delta\phi_{NL}(\tau{=}15\,\text{ ms})$ and $\theta_i$ is shown in Fig. \ref{fig:results}b. The black curve is a Gaussian fit. The mean value is determined to be $5 \pm 5.4 \times 10^{-12}$ where the reported uncertainty corresponds to 1 standard deviation. The average uncertainty of the individual measurements computed using standard propagation of error when solving the system of equations Eqs. \eqref{stdphi}, \eqref{deltaphinl} and assuming only QPN is found to be $7.7 \times 10^{-11}$, which is in good agreement with the sample standard deviation $8.2 \times 10^{-11}$. 

In summary, we have improved the bound of potential nonlinearities of a causal extension to QM from $\tilde{\epsilon}_{\gamma}\lessapprox 10^{-2}$ to $|\tilde{\epsilon}_{\gamma}| \lessapprox 5.4 \times 10^{-12}$.  Further improvements could be achieved with longer averaging times, longer coherence times, or sophisticated quantum measurement protocols such as using squeezed states. Similarly tighter bounds can also be achieved by localizing the test particle better, for instance, by increasing the mass $m$ or the confinement $\nu$. 

During the preparation of this manuscript, we became aware of related work \cite{Polkovnikov2022Apr}.

\begin{acknowledgments}
 J.B., S.K. and H.H. acknowledge funding by the U.S. Department of Energy, Office of Science, Office of Basic Energy Sciences under Award Number DE{-}SC0019376.
\end{acknowledgments}

\bibliography{bib}

\end{document}